\documentclass[twoside,fleqn]{article}
\usepackage{npb}
\usepackage{graphicx}
\newcommand{\AmS}{{\protect\the\textfont2
  A\kern-.1667em\lower.5ex\hbox{M}\kern-.125emS}}
\newenvironment{Eqnarray}%
         {\arraycolsep 0.14em\begin{eqnarray}}{\end{eqnarray}}
\def\eq#1{eq.~(\ref{#1})}
\def\eqs#1#2{eqs.~(\ref{#1})--(\ref{#2})}
\def\Eq#1{Eq.~(\ref{#1})}

\def\eqns#1#2{eqs.~(\ref{#1}) and (\ref{#2})}

\def\fig#1{fig.~\ref{#1}}
\def\beqa{\begin{Eqnarray}}
\def\eeqa{\end{Eqnarray}}
\def\beq{\begin{equation}}
\def\eeq{\end{equation}}
\def\beqno{\begin{Eqnarray}}
\def\eeqno{\end{Eqnarray}}
\def\crr{\crcr\noalign{\vskip .1in}}
\def\ifmath#1{\relax\ifmmode #1\else $#1$\fi}
\def\ls#1{\ifmath{_{\lower1.5pt\hbox{$\scriptstyle #1$}}}}
\def\half{\ifmath{{\textstyle{1 \over 2}}}}
\def\nicefrac#1#2{\hbox{${#1\over #2}$}}
\def\tanb{\tan\beta}
\def\cotb{\cot\beta}
\def\mz{m_Z}
\def\mw{m_W}
\def\hsm{h\ls{\rm SM}}

\def\hh{H}
\def\hl{h}
\def\ha{A}

\def\mhh{m_{H}}
\def\mhl{m_{h}}
\def\mha{m_{A}}
\def\mhpm{m_{H^\pm}}
\def\lsim{\mathrel{\raise.3ex\hbox{$<$\kern-.75em\lower1ex\hbox{$\sim$}}}}
\def\gsim{\mathrel{\raise.3ex\hbox{$>$\kern-.75em\lower1ex\hbox{$\sim$}}}}
\def\lam{\lambda}
\def\lambar{\lam}

\def\lamhat{\widehat\lam}
\def\sb  {s_{\beta}}
\def\cb  {c_{\beta}}
\def\stwob  {s_{2\beta}}
\def\ctwob  {c_{2\beta}}
\def\sa  {s_{\alpha}}
\def\ca  {c_{\alpha}}

\def\ctwob{c_{2\beta}}

\def\stwob{s_{2\beta}}

\def\cbma{c_{\beta-\alpha}}

\def\cthreeb{c_{3\beta}}

\def\sthreeb{s_{3\beta}}

\begin{document}

\topmargin-2pc
\vbox{  \large
\begin{flushright}
SCIPP 02/27 \\
November, 2002 \\
hep--ph/0212010\\
\end{flushright}
\vskip1.5cm
\begin{center}
{\LARGE\bf
Decoupling and the Radiatively-Corrected  \\[6pt]
MSSM Higgs Sector}\\[1cm]

{\Large Howard E. Haber}\\[5pt]{\it Santa Cruz Institute for Particle
Physics,  \\ University of California, Santa Cruz, CA 95064}\\

\vskip2cm
\thispagestyle{empty}

{\bf Abstract}\\[1pc]

\begin{minipage}{15cm}
In the decoupling limit of a non-minimal Higgs sector, the lightest
CP-even Higgs boson ($\hl$) is indistinguishable from the Standard
Model (SM) Higgs boson.  In the two-Higgs-doublet sector of the
MSSM, the approach to the decoupling limit
(for $\mha\gg\mz$) persists, even in the presence of potentially
large ($\tanb$-enhanced) radiative corrections to the $\hl b\bar b$
coupling.  Radiative corrections can also generate an accidental
cancellation between tree-level and one-loop terms, resulting in
a SM-like Higgs boson for moderate $\mha$ outside the decoupling regime.
\end{minipage}  \\
\vskip2.5cm
Invited talk at the \\
6th International Symposium on Radiative Corrections: \\
Application of Quantum Field Theory to Phenomenology---RADCOR 2002\\
and 6th Zeuthen Workshop on Elementary Particle Theory: \\
Loops and Legs in Quantum Field Theory \\
September 8--13, 2002, Kloster Banz, Germany\\
\end{center}
}
\vfill
\clearpage

\setcounter{page}{1}

\title{Decoupling and the radiatively-corrected MSSM Higgs sector
        \thanks{This work was supported in part by the U.S.
         Department of Energy under the grant DE-FG03-92ER40689.}}

\author{Howard E.~Haber\address{Santa Cruz Institute for Particle Physics,\\
        University of California, Santa Cruz, CA 95064 USA}}%


\begin{abstract}
In the decoupling limit of a non-minimal Higgs sector, the lightest
CP-even Higgs boson ($\hl$) is indistinguishable from the Standard
Model (SM) Higgs boson.  In the two-Higgs-doublet sector of the
MSSM, the approach to the decoupling limit
(for $\mha\gg\mz$) persists, even in the presence of potentially
large ($\tanb$-enhanced) radiative corrections to the $\hl b\bar b$
coupling.  Radiative corrections can also generate an accidental
cancellation between tree-level and one-loop terms, resulting in
a SM-like Higgs boson for moderate $\mha$ outside the decoupling regime.
\end{abstract}

\maketitle

\section{Introduction}
Suppose a Higgs boson is discovered at the LHC, and its properties are
observed to coincide (within experimental error) to those of the
Standard Model (SM) Higgs boson.  Moreover, suppose that evidence for
supersymmetry is found, which suggests that in the minimal
version of the model (MSSM), the observed Higgs boson is the lightest
state of a two-Higgs-doublet model (2HDM).  Finally, imagine that 
no evidence for the heavier Higgs states at the LHC is found. 
With precision measurements at a future high energy
$e^+e^-$ Linear Collider (LC), can one determine the mass scale of the
heavier Higgs states?

At the LHC, more than one Higgs scalar of the MSSM will often be observed if
$\tan\beta\gg 1$ (due to enhanced Higgs couplings to down-type fermions).
But, there is also a substantial region of moderate $\tanb$
in which only the lightest CP-even Higgs scalar ($\hl$) is 
observed~\cite{howmar}.  If the properties of $\hl$ approximate
those of the SM Higgs boson ($\hsm$), a program of precision Higgs
measurements at the LC will play a critical role
in elucidating the physics of Higgs bosons.

\section{Decoupling Limit of the 2HDM~\cite{gunion}}

Given a non-minimal Higgs sector, the decoupling limit corresponds to
the parameter regime in which all but one CP-even neutral Higgs scalar are
significantly heavier than the $Z$.  The properties of the lightest
CP-even Higgs boson are nearly indistinguishable from those of the
Standard Model (SM) Higgs boson. The decoupling limit is very
general and exists in many multi-Higgs models.
The MSSM Higgs sector provides a well-motivated example for the
decoupling limit and is the main focus of this work.

First, consider the general 2HDM.  The most
general scalar potential is given by:
\beqno
&&\mathcal{V}= m_{11}^2\Phi_1^\dagger\Phi_1+m_{22}^2\Phi_2^\dagger\Phi_2
-[m_{12}^2\Phi_1^\dagger\Phi_2+{\rm h.c.}]\nonumber \\
&&+\half\lambda_1(\Phi_1^\dagger\Phi_1)^2
+\half\lambda_2(\Phi_2^\dagger\Phi_2)^2
+\lambda_3(\Phi_1^\dagger\Phi_1)(\Phi_2^\dagger\Phi_2) \nonumber \\
&&+\lambda_4(\Phi_1^\dagger\Phi_2)(\Phi_2^\dagger\Phi_1)
+\half\left[\lambda_5(\Phi_1^\dagger\Phi_2)^2+{\rm
h.c.}\right]\nonumber \\
&&+\left\{\left[\lambda_6(\Phi_1^\dagger\Phi_1)
+\lambda_7(\Phi_2^\dagger\Phi_2)\right]
\Phi_1^\dagger\Phi_2+{\rm h.c.}\right\}\,.
\eeqno 
For simplicity, we assume that there is no explicit (or
spontaneous) CP violation.  Since the ground state must preserve
U(1)$_{\rm EM}$, the scalar vacuum expectation values are
$\langle \Phi_i^0 \rangle\equiv v_i/\sqrt{2}$, with $\tan\beta\equiv
v_2/v_1$ and $v^2\equiv v_1^2+v_2^2=(246~{\rm GeV})^2$.  

Diagonalizing the CP-even Higgs squared-mass matrix yields
two CP-even scalar eigenstates:
\beqno
\hl &=&-(\sqrt{2}{\rm Re\,}\Phi_1^0-v_1)\sa+
(\sqrt{2}{\rm Re\,}\Phi_2^0-v_2)\ca\,,\nonumber\\
\hh &=&(\sqrt{2}{\rm Re\,}\Phi_1^0-v_1)\ca+
(\sqrt{2}{\rm Re\,}\Phi_2^0-v_2)\sa\,,
\eeqno
where $\ca\equiv\cos\alpha$, $\sa\equiv\sin\alpha$.
The other Higgs scalars of the model include
a CP-odd state, $A$, and a charged
scalar pair, $H^\pm$.  The decoupling limit is defined as the limit of
$\mha\gg m_Z$, assuming $\lambda_i\lsim\mathcal{O}(1)$.  One can show
that this limit corresponds to taking $\beta-\alpha\to\pi/2$; {\it
i.e.}, $\cos(\beta-\alpha)\to 0$.  In the approach to the
decoupling limit, one finds~\cite{gunion}:
\beqno 
&&\mha^2 \,\,\simeq \,\,
v^2 \left[{\lamhat\over\cbma}+\lam_A-\nicefrac{3}{2}\lamhat\,\cbma
\right]\,,\label{decoupmasses1}\\
&&\mhl^2 \,\,\simeq \,\, v^2(\lambar-\lamhat\,\cbma)\,,\label{decoupmasses2}\\
&&\mhh^2 \,\,\simeq \,\, \mha^2
+(\lambda-\lambda_A+\lamhat\,\cbma)v^2\,,\label{decoupmasses3}\\
&&\mhpm^2\,\,= \,\, \mha^2+\half(\lam_5-\lam_4)v^2\,.\label{decoupmasses4}
\eeqno
where $\cbma\equiv\cos(\beta-\alpha)$ and
\beqno
&&\lambar \equiv \lam_1\cb^4+\lam_2\sb^4
+\half(\lam_3+\lam_4+\lam_5)\stwob^2 \nonumber \\
&&\qquad\qquad +2\stwob(\lam_6\cb^2+\lam_7\sb^2)
\,, \\
&&\lamhat \equiv \half\stwob\left[\lam_1\cb^2
-\lam_2\sb^2-(\lam_3+\lam_4+\lam_5)\ctwob\right]\nonumber \\
&&\qquad\qquad -\lam_6\cb\cthreeb-\lam_7\sb\sthreeb
\,,\\
&&\lam_{A}\equiv
\ctwob(\lam_1\cb^2-\lam_2\sb^2)+(\lam_3+\lam_4) s_{2\beta}^2
\nonumber \\
&&\qquad\qquad -\lam_5 c_{2\beta}^2+2\lam_6\cb\sthreeb-2\lam_7\sb\cthreeb\,.
\eeqno
In particular, \eqs{decoupmasses1}{decoupmasses3} yield
\beq \label{cbmadecoup}
\cos(\beta-\alpha)\simeq  {\lamhat v^2 \over \mha^2-\lambda_A v^2}
\simeq{\lamhat v^2\over \mhh^2-\mhl^2}\,.
\eeq
It follow that: (i) $\mhl\sim \mathcal{O}(m_Z)$,
(ii) $\mhh\simeq  \mha\simeq\mhpm$, up to corrections
of $\mathcal{O}(m_Z^2/\mha)$, and (iii)
$\cos(\beta-\alpha)\sim \mathcal{O}(m_Z^2/\mha^2)$.

The couplings of the Higgs bosons to vector bosons, fermions and
scalars typically depend on $\alpha$ and $\beta$.  By examining the
tree-level couplings of the lightest CP-even scalar $\hl$, one notes
that in the limit of $\cbma=0$, the couplings of $\hl$ reduce to the
corresponding SM Higgs couplings.  That is, in the approach to the 
decoupling limit, the properties of $\hl$ are nearly indistinguishable
from those of the SM Higgs boson, whereas all the other Higgs states are
significantly heavier.  Thus, the effective low energy theory
below the mass scale of $\mathcal{O}(\mha)$ is the
Standard Model with one Higgs doublet.

The Higgs sector of the minimal supersymmetric extension of the
Standard Model (MSSM) is a two-Higgs doublet model, with interactions
constrained by supersymmetry~\cite{susyhiggs}.  It particular, 
\beqa \label{susylams}
&&\lam v^2=\mz^2 \ctwob^2\,,\qquad \quad\,\;\lam_4 v^2=-2\mw^2\,,\nonumber \\
&&\lamhat v^2=\mz^2\stwob\ctwob\,,\qquad \lam_A v^2=\mz^2 c_{4\beta}\,,
\eeqa
and $\lam_5=\lam_6=\lam_7=0$.
Moreover, the following tree-level result can be derived:
\beq \label{cbmaexact}
\cos^2(\beta-\alpha)={\mhl^2(\mz^2-\mhl^2)\over
\mha^2(\mhh^2-\mhl^2)}\,.
\eeq
As expected, $\cbma\to 0$ in the decoupling limit where 
$\mhl\sim\mathcal{O}(m_Z)$ and $\mha\gg m_Z$. 
Moreover, when $m_A\gg m_Z$, \eqs{decoupmasses1}{decoupmasses4} yield
\beqa \label{decoupmasses}
&&\mhl^2 \simeq\mz^2\ctwob^2\,,\qquad
\mhh^2 \simeq\mha^2+\mz^2\stwob^2\,,\nonumber \\
&&\mhpm^2 = \mha^2+\mw^2\,,\quad 
\cbma\simeq {\mz^4\sin^2 4\beta\over
4\mha^4}\,.
\eeqa

\section{A SM-like Higgs boson without decoupling~\cite{gunion}}

It is possible for the theory to exhibit a SM-like Higgs
boson without decoupling.  For example, assume that $\tanb\geq 1$.  Two
cases arise in the 2HDM
where $|\cbma|\ll 1$: (i) $\mha^2\gg \lam_i
v^2\tanb$, with $\lambda_i\lsim\mathcal{O}(1)$, and (ii) $|\lamhat|\ll
1$ with $\mha$ arbitrary.  
(If $\tanb\leq 1$, replace $\tanb$ with $\cotb$ above.)
In the MSSM, the $\hl b\bar b$ coupling
normalized to its SM value is given by:
\beq
-{\sin\alpha\over\cos\beta}=\sin(\beta-\alpha)
-\tan\beta\,\cos(\beta-\alpha)\,.
\eeq  
In case (i), we have $|\tanb\,\cbma|\ll 1$ even if $\tanb\gg 1$ 
[see \eq{cbmadecoup}], which
implies that the $\hl b\bar b$ coupling is SM-like 
(corresponding to the decoupling limit).  
Case (ii) does not correspond to decoupling if
$\mha^2\lsim\mathcal{O}(v^2)$.
If $\tanb\gg 1$, it is possible to have
$|\tanb\,\cbma|\sim\mathcal{O}(1)$ even if $|\cbma|\ll 1$, in which case
the $\hl b\bar b$ coupling deviates from its SM value.
Nevertheless, for the Higgs couplings to $t\bar t$, vector bosons and scalars,
$\hl$ is SM-like. 

In the MSSM at tree-level, $|\cbma|\ll 1$ is possible
only in the decoupling regime [see \eq{cbmaexact}], 
corresponding to case (i) above.
However, one-loop effects mediated by 
supersymmetric particles can generate significant modifications to the
tree-level MSSM Higgs sector.  For example, the tree-level upper bound,
$\mhl\leq\mz$, can be significantly raised~\cite{hhprl}.  Allowing for maximal
mixing in the top squark sector and supersymmetric mass parameters of
order $M_S\sim 1$~TeV, one finds a radiatively-corrected Higgs mass bound of
$\mhl\lsim 135$~GeV~\cite{howmar}.  We shall demonstrate in section 4 that
for moderate values of $\mha$, radiative corrections to $\lamhat$ 
can result in $|\lamhat|\ll 1$ in certain regions of the MSSM parameter
space.  This would correspond to case (ii) above, and
the $\hl$ of the MSSM could indeed exhibit SM-like properties
outside the domain of the decoupling limit.

\section{Radiatively-corrected MSSM Higgs couplings~\cite{chlm}}

In order to study the decoupling properties of the MSSM Higgs sector,
it is crucial to examine the Higgs couplings, including
the most significant loop-corrections.
The leading contributions to the radiatively-corrected Higgs
couplings arise in two ways.  First, the radiative corrections
to the CP-even Higgs squared-mass matrix results in a shift 
of the CP-even Higgs mixing angle $\alpha$ from its tree-level value. 
That is, the dominant Higgs propagator corrections 
can to a good approximation be absorbed into an effective 
(``radiatively-corrected'') mixing angle $\alpha$~\cite{hffsusyprop}.
In this approximation, we can write:
\beq \label{calmatrix} 
{\cal M}^2\equiv
\left( \matrix{{\cal M}_{11}^2 &  {\cal M}_{12}^2 \crr {\cal
M}_{12}^2 &  {\cal M}_{22}^2 } \right) ={\cal M}_0^2+\delta {\cal
M}^2\,, 
\eeq 
where the tree-level contribution is denoted by ${\cal M}_0^2$ 
and $\delta {\cal M}^2$ is the contribution from
the radiative corrections.
Then, $\cbma$ is given by
\beq \label{eq:cosbma}
\cbma={(\mathcal{M}_{11}^2-\mathcal{M}_{22}^2)\sin 2\beta
-2\mathcal{M}_{12}^2\cos 2\beta\over
2(m_H^2-m_h^2)\sin(\beta-\alpha)} \,.
\eeq
Inserting the tree-level values for $\mathcal{M}^2_0$, one can rewrite
\eq{eq:cosbma} as
\beqa \label{eq:cosbmarad}
&&\cbma={m_Z^2\sin 4\beta\over
  2(m_H^2-m_h^2)\sin(\beta-\alpha)}\nonumber \\[4pt]
&&+{({\delta\mathcal{M}}_{11}^2-{\delta\mathcal{M}}_{22}^2)\sin 2\beta
-2{\delta\mathcal{M}}_{12}^2\cos 2\beta\over
2(m_H^2-m_h^2)\sin(\beta-\alpha)}\,.
\eeqa
Using tree-level Higgs couplings with $\alpha$ replaced by its 
effective one-loop value provides a useful first approximation to 
the radiatively-corrected Higgs couplings.

Second, contributions from the
one-loop vertex corrections to tree-level Higgs-fermion couplings can
modify these couplings in a significant way, especially
in the limit of large $\tan\beta$.  When
radiative corrections are included, all possible dimension-four
Higgs-fermion couplings are generated.  In particular, the effects of
higher dimension operators can be ignored if $M_S\gg\mz$, which we
henceforth assume.
These results can be summarized by an effective
Lagrangian that describes the coupling of
the neutral Higgs bosons to the third generation
quarks:
\beqa \label{susyyuklag}
        -\mathcal{L}_{\rm eff} &=& 
        (h_b + \delta h_b) \bar b_R b_L \Phi_1^{0\ast} 
        + (h_t + \delta h_t) \bar t_R t_L \Phi_2^0 \nonumber \\
      &  +& \Delta h_t \bar t_R t_L \Phi_1^0 
        + \Delta h_b \bar b_R b_L \Phi_2^{0 \ast}
        + {\rm h.c.}\,,
\eeqa
resulting in a modification of the tree-level relation between
$h_t$ [$h_b$] and $m_t$ [$m_b$] as
follows~\cite{deltamb,hffsusyqcd,deltamb2}:
\beqa
        m_b &=& \frac{h_b v}{\sqrt{2}} \cos\beta
        \left(1 + \frac{\delta h_b}{h_b}
        + \frac{\Delta h_b \tan\beta}{h_b} \right) \nonumber \\
        &&\qquad\equiv\frac{h_b v}{\sqrt{2}} \cos\beta
        (1 + \Delta_b)\,, \label{byukmassrel} \\[5pt]
        m_t &=& \frac{h_t v}{\sqrt{2}} \sin\beta
        \left(1 + \frac{\delta h_t}{h_t} + \frac{\Delta
        h_t\cot\beta}{h_t} \right)\nonumber \\
        &&\qquad \equiv\frac{h_t v}{\sqrt{2}} \sin\beta
        (1 + \Delta_t)\,. \label{tyukmassrel}
\eeqa
The dominant contributions to $\Delta_b$ are $\tan\beta$-enhanced,
with $\Delta_b\simeq (\Delta h_b/h_b)\tan\beta$; whereas for
$\tan\beta\gg 1$, $\delta h_b/h_b$ provides a small correction to
$\Delta_b$.   [In the same limit, $\Delta_t\simeq\delta h_t/h_t$, with
the additional contribution of $(\Delta h_t/h_t)\cot\beta$ providing a
small correction.]

From \eq{susyyuklag} we can obtain the couplings of the physical 
neutral Higgs
bosons to third generation quarks.  The resulting couplings of $\hl$
to $b\bar b$ and $t\bar t$ pairs are given by:
\beqa
 g_{\hl t\bar t} & = & {m_t\over v}{\cos\alpha \over \sin\beta}
\left[1-{1\over 1+\Delta_t}{\Delta h_t\over h_t}
(\cot\beta+ \tan\alpha)\right]\nonumber \\[4pt]
 g_{\hl b\bar b} &= & -{m_b\over v}{\sin\alpha \over \cos\beta}
\left[1+{1\over 1+\Delta_b}\right.\nonumber \\
&&\quad\;\times \left.\left({\delta h_b\over h_b}-
\Delta_b\right)\left( 1 +\cot\alpha \cot\beta \right)\right]\,.
\label{hlff}
\eeqa

We now turn to the decoupling limit.  First
consider the implications for the radiatively-corrected value of
$\cbma$.  
Since $\delta\mathcal{M}^2_{ij}\sim {\mathcal O}(m_Z^2)$,
and $m_H^2-m_h^2=m_A^2+\mathcal{O}(m_Z^2)$, one finds~\cite{chlm}
\begin{equation} \label{cosbmadecoupling}
        \cos(\beta-\alpha)=c\left[{m_Z^2\sin 4\beta\over
        2m_A^2}+\mathcal{O}\left(m_Z^4\over m_A^4\right)\right]\,,
\end{equation}
in the limit of $\mha\gg\mz$, where
\begin{equation} \label{cdef}
        c\equiv 1+{{\delta\mathcal{M}}_{11}^2-{\delta\mathcal{M}}_{22}^2\over
        2m_Z^2\cos 2\beta}-{{\delta\mathcal{M}}_{12}^2\over m_Z^2\sin
        2\beta}\,.
\end{equation}
Equivalently, the radiative corrections have modified the
tree-level definition of $\lamhat$:
\beq \label{lamhatrad}
\lamhat v^2= c\mz^2\sin 2\beta\cos 2\beta\,.
\eeq
\Eq{cosbmadecoupling} exhibits the expected decoupling behavior
for $m_A\gg m_Z$.
However, \eqns{eq:cosbmarad}{cosbmadecoupling} exhibit another way in which
$\cos(\beta-\alpha)=0$ can be achieved---simply choose the
MSSM parameters (which govern the Higgs mass radiative
corrections) such that the right hand side of \eq{eq:cosbmarad} vanishes.
That is,
\begin{equation}
        \sin 2\beta =
        {2\, \delta \mathcal{M}^2_{12}
        - \tan 2\beta
        \left(\delta \mathcal{M}^2_{11} - \delta \mathcal{M}^2_{22}\right) 
        \over 2\mz^2}\,.
        \label{eq:tanbetadecoup}
\end{equation}
This condition is equivalent to $c=0$, which implies $\lamhat=0$ 
[\eq{lamhatrad}] and thus corresponds to case (ii) in section 3.  That is,
although $\lamhat\neq 0$ at tree-level, the
one-loop radiative corrections to $\lamhat$ can approximately cancel its
tree-level value, resulting in $|\lamhat|\ll 1$.  
(Note that the one-loop corrections arise from the
exchange of supersymmetric particles, whose contributions can be
enhanced for certain MSSM parameter choices.  
One can show that the 
two-loop corrections are subdominant, so that the approximation scheme
is under control.)  In particular,
\eq{eq:tanbetadecoup} is independent of the value of $m_A$.
Typically,
\eq{eq:tanbetadecoup} yields a solution at large $\tan\beta$.  That
is, by
approximating $\tan 2\beta\simeq -\sin 2\beta \simeq -2/ \tan \beta$,
one can determine
the value of $\beta$ at which $\lamhat\simeq 0$~\cite{chlm}:
\begin{equation} \label{earlydecoupling}
\tan \beta\simeq \frac{2m_Z^2-
\delta\mathcal{M}_{11}^2+\delta \mathcal{M}_{22}^2}
{ \delta\mathcal{M}_{12}^2}\,.
\end{equation}
Hence, there exists a value of $\tanb$ (which depends on the choice
of MSSM parameters) where
$\cos(\beta-\alpha)\simeq 0$ independently of
the value of $m_A$.  If $\mha$ is not much larger than
$\mz$, then $\hl$ is
a SM-like Higgs boson outside the decoupling regime.

Finally, note that for $\mha\gg\mz$,
\begin{equation}  \label{cotalf}
\cot\alpha = -\tan\beta\left[1+ \frac{2\mz^2}{\mha^2}\ctwob\right] 
+ {\cal O}\left(\frac{\mz^4}{\mha^4}\right)\,.
\end{equation}
Applying this result to \eq{hlff}, it follows 
that in the decoupling limit, $g_{\hl q\bar q}=
g_{\hsm q\bar q}=m_q/v$.
Away from the decoupling limit,
the Higgs couplings to down-type fermions can deviate significantly
from their tree-level values due to enhanced radiative corrections at
large $\tan\beta$ [where $\Delta_b\simeq\mathcal{O}(1)$].
In particular, because $\Delta_b\propto\tan\beta$, the leading
one-loop radiative correction to $g_{\hl b\bar b}$ is of
$\mathcal{O}(\mz^2\tanb/\mha^2)$, which decouples only when
$\mha^2\gg\mz^2\tanb$ (this behavior was called {\it delayed
decoupling} in \cite{loganetal}).
  
\section{Implications for precision Higgs measurements at the LC}

\begin{figure}[!ht]
\begin{center}
\resizebox{0.45\textwidth}{!}{
\includegraphics*[19,142][529,682]{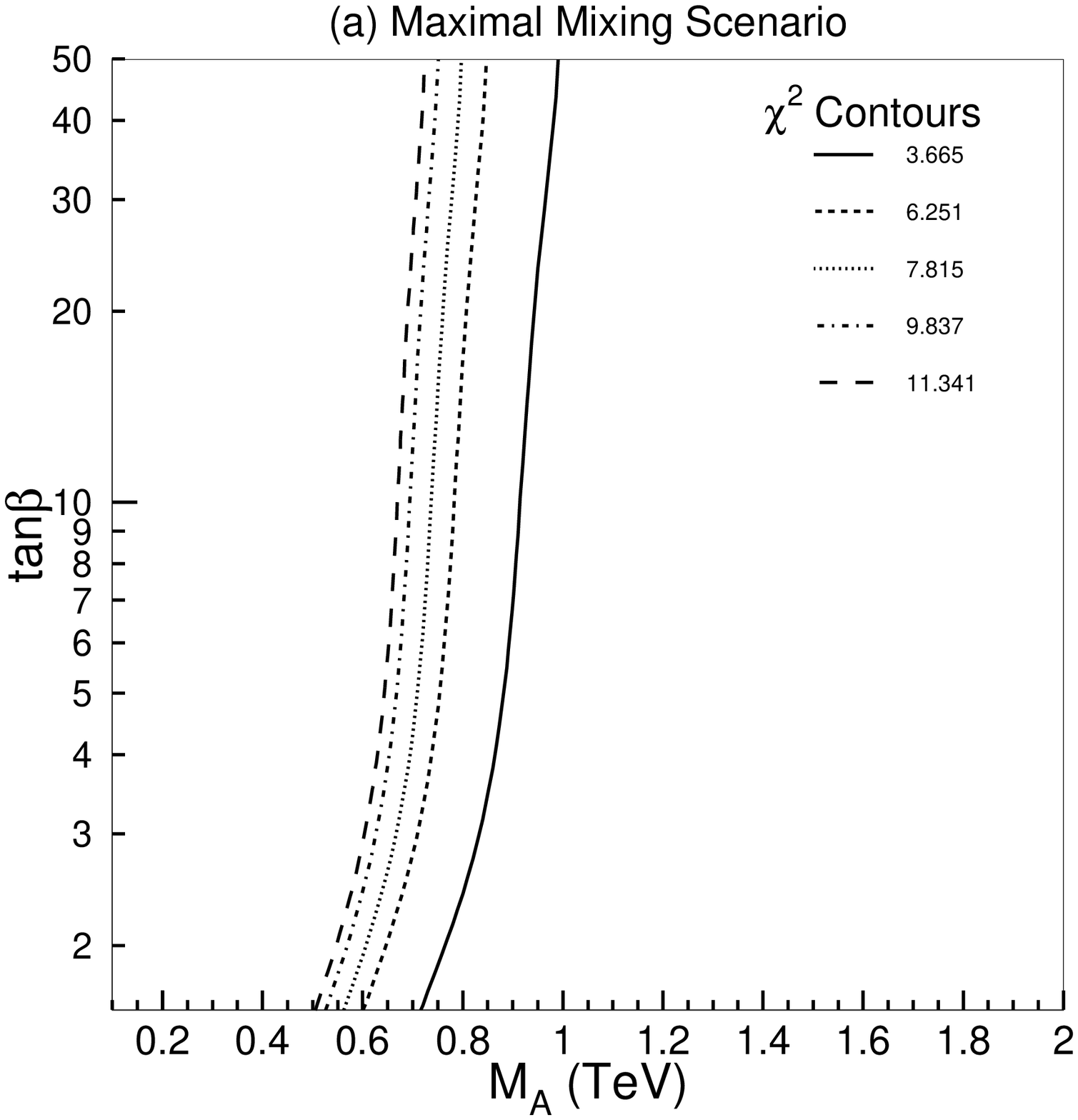}}
\resizebox{0.45\textwidth}{!}{
\includegraphics*[19,142][529,682]{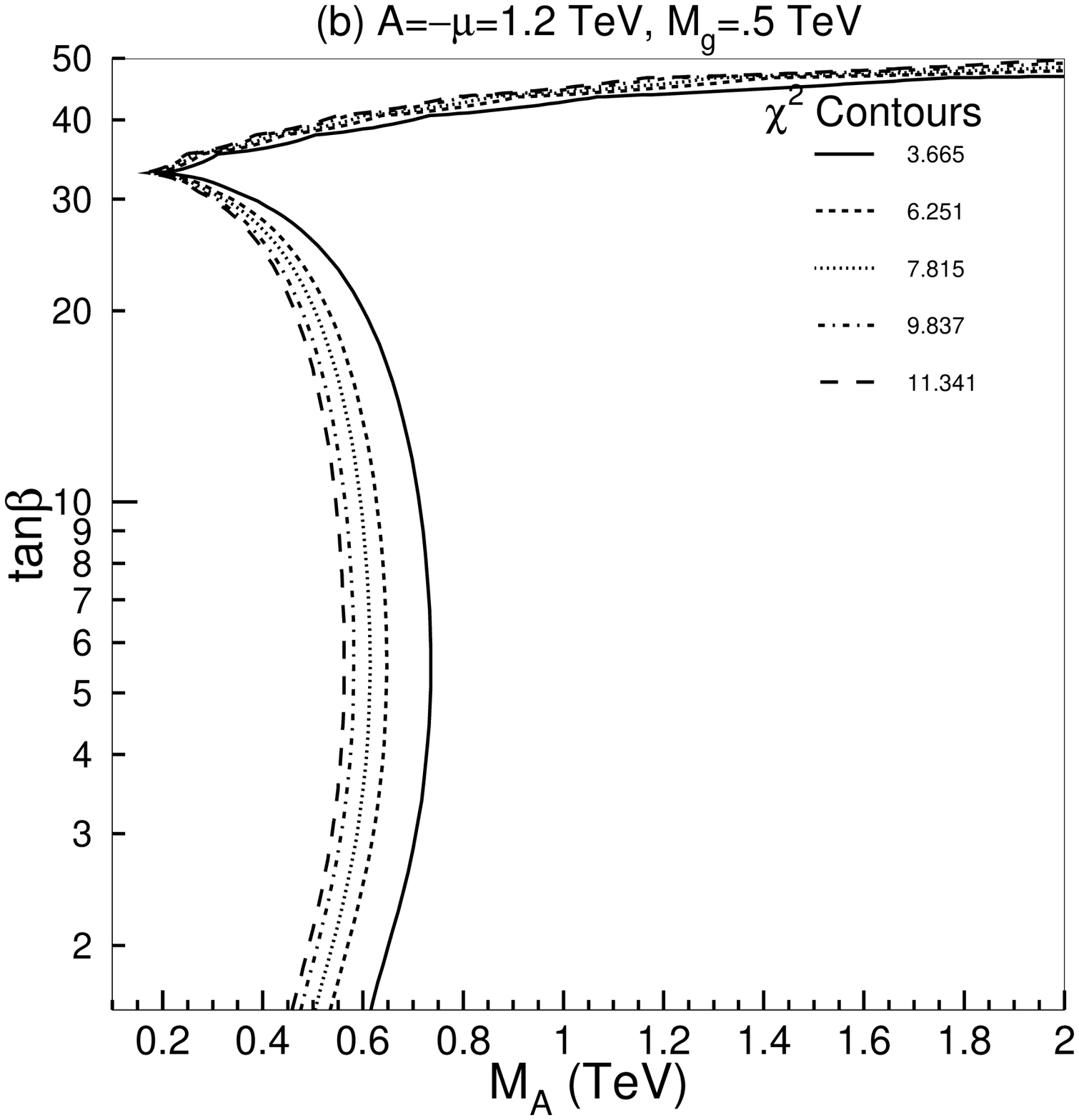}
}
\end{center}
\vspace{-0.5in}
\caption{ Contours of $\chi^2$ for Higgs
boson decay observables for (a) the maximal mixing scenario; and
(b) a choice of MSSM parameters for which the loop-corrected
$\hl b\bar b$ coupling is suppressed at large $\tanb$ and low $\mha$
(relative to the corresponding tree-level coupling).
The contours correspond to
68, 90, 95, 98 and 99\% confidence levels (right to left) for the 
observables $g^2_{hbb}$, $g^2_{h\tau\tau}$, and $g^2_{hgg}$. See
\cite{chlm} for additional details.}
\label{fig:chisquare}
\end{figure}

As noted in section 1, a program of
precision Higgs measurements at the LC may be critical in
determining whether the properties of the lightest CP-even Higgs boson
differ from those of $\hsm$.  In particular, recent simulations
of Higgs branching ratio 
measurements~\cite{battaglia} suggest that the Higgs couplings
to vector bosons and the third generation fermions can be determined with
an accuracy in the range of 1--3\% at the LC.  
In the exact decoupling limit (of infinitely large $\mha$),
$\hl=\hsm$.  However, for finite values of $\mha$,
the fractional deviations of the couplings of $\hl$ relative
to those of $\hsm$ scale as $m_Z^2/m_A^2$.  Thus,
if precision measurements reveal
a significant deviation from SM expectations, 
one could in principle derive a constraint 
({\it e.g.}, upper and lower bounds) on the
heavy Higgs masses of the model. 

In the MSSM, this constraint is
sensitive to the supersymmetric parameters that control the radiative
corrections to the Higgs couplings.  This is illustrated in
\fig{fig:chisquare}, where the constraints on $\mha$ are derived
for two different sets of MSSM parameter 
choices~\cite{chlm}.  Here, a simulation of a
global fit of measured $hbb$, $h\tau\tau$ and $hgg$ couplings is made
(based on the anticipated experimental accuracies given in \cite{battaglia})
and $\chi^2$ contours are plotted indicating the constraints in the
$m_A$--$\tan\beta$ plane, assuming that a deviation from SM Higgs
boson couplings is seen.
In the maximal mixing scenario shown in \fig{fig:chisquare}(a),
the constraints on $\mha$ are significant and rather insensitive to the
value of $\tan\beta$. 
However in some cases, as shown in \fig{fig:chisquare}(b),
a region of $\tan\beta$ may yield almost no constraint on $\mha$.
This corresponds to the value of $\tanb$ given by
\eq{earlydecoupling}, and is a result of $\lamhat\simeq 0$ generated by
radiative corrections [$c\simeq 0$ in \eq{lamhatrad}].
Thus, one cannot extract a fully model-independent upper bound on
the value of $\mha$ beyond the kinematical limit that would be obtained
if direct $\ha$ production were not observed at the LC.

%



\begin{thebibliography}{9}

 
\bibitem{howmar}
For a review and references, see M.~Carena and H.E.~Haber,
FERMILAB-Pub-02/114-T and SCIPP 02/07
[hep-ph/0208209].


\bibitem{gunion}
J.F.~Gunion and H.E.~Haber, UCD-2002-10 and SCIPP-02/10
[hep-ph/0207010].

\bibitem{susyhiggs}
K.~Inoue, A.~Kakuto, H.~Komatsu, and S.~Takeshita,
Prog.\ Theor.\ Phys.\ {\bf 67} (1982) 1889; 
R.~Flores and M.~Sher, Annals Phys.\ {\bf 148} (1983) 95;
J.F.~Gunion and H.E.~Haber, Nucl.\ Phys. {B272} (1986) 1 
[E: {\bf B402} (1993) 567].

\bibitem{hhprl}
H.E.~Haber and R.~Hempfling,
Phys.\ Rev.\ Lett.\  {\bf 66} (1991) 1815;
Y.~Okada, M.~Yamaguchi and T.~Yanagida,  
Prog.\ Theor.\ Phys.\  {\bf 85} (1991) 1;
J.R.~Ellis, G.~Ridolfi and F.~Zwirner,
Phys.\ Lett.\ {\bf B257} (1991) 83.


\bibitem{hffsusyprop}
S.~Heinemeyer, W.~Hollik and G.~Weiglein,
Eur.\ Phys.\ J.\ {\bf C16} (2000) 139.

\bibitem{deltamb}
R.~Hempfling, Phys.\ Rev.\ {\bf D49} (1994) 6168;
L.J.~Hall, R.~Rattazzi and U.~Sarid,  
Phys.\ Rev.\ {\bf D50} (1994) 7048;
M.~Carena, M.~Olechowski, S.~Pokorski and C.E.M.~Wagner, 
Nucl.\ Phys.\ {\bf B426} (1994) 269.

\bibitem{hffsusyqcd}
A.~Bartl, {\it et al.}, 
Phys.\ Lett.\ {\bf B378} (1996) 167;
R.A.~Jim{\'e}nez and J.~Sol{\`a},           
Phys.\ Lett.\ {\bf B389} (1996) 53;
J.A.~Coarasa, R.A.~Jim\'enez and J.~Sol\`a, 
Phys.\ Lett.\  {\bf B389} (1996) 312.


\bibitem{deltamb2}
D.M.~Pierce, J.A.~Bagger, K.~Matchev, and R.~Zhang, 
Nucl.\ Phys.\ {\bf B491} (1997) 3.

\bibitem{chlm}
M.~Carena, H.E.~Haber, H.E.~Logan and S.~Mrenna,
Phys.\ Rev.\ {\bf D65} (2002) 055005 
[E:\ {\bf D65} (2002) 099902].

\bibitem{loganetal}
H.E. Haber, M.J. Herrero, H.E. Logan, S. Pe\~naranda, S. Rigolin
and D. Temes, Phys.\ Rev.\ {\bf D63} (2001) 055004.

\bibitem{battaglia}
M.~Battaglia and K.~Desch, hep-ph/0101165.



\end{thebibliography}
\end{document}